\documentclass[showpacs,amsmath,amssymb,superscriptaddress,twocolumn,prd,aps,10pt]{revtex4-1}
\input{tcilatex}
\usepackage{graphicx,amsmath,color}
\usepackage{graphicx}
\usepackage{psfrag}
\newcommand{\ve}{\varepsilon}

\begin{document}
\title{Thick brane solutions and topology change transition on black hole backgrounds}
\author{Viktor G.~Czinner}
\affiliation{Department of Mathematics and Applied Mathematics,\\
University of Cape Town, 7701 Rondebosch, Cape Town, South Africa;}
\email{czinner@rmki.kfki.hu}
\affiliation{Department of Theoretical Physics,\\
MTA KFKI Research Institute for Particle and Nuclear Physics,\\ 
Budapest 114, P.O.~Box 49, H-1525, Hungary}

\begin{abstract}
We consider static, axisymmetric, thick brane solutions on higher dimensional, spherically symmetric 
black hole backgrounds. It was found recently \cite{CF}, that in cases when the thick brane has more 
than 2 spacelike dimensions, perturbative approaches break down around the corresponding thin solutions 
for Minkowski type topologies. This behavior is a consequence of the fact that thin solutions are 
not smooth at the axis, and for a general discussion of possible phase transitions in the 
system, one needs to use a non-perturbative approach. In the present paper we provide an exact, numerical 
solution of the problem both for black hole- and Minkowski type topologies with arbitrary number of brane 
and bulk dimensions. We also illustrate a topology change transition in the system for 
a 5-dimensional brane embedded in a 6-dimensional bulk.

\end{abstract}

\maketitle

\section{Introduction}

Curvature corrections to Dirac-Nambu-Goto (DNG) membranes \cite{Dirac, Nambu, Goto} in higher dimensional black hole 
backgrounds attracts a lot of attention recently in several different areas of modern physics. The simplest 
corrections to the DNG brane action originate from small thickness perturbations that are quadratic at the 
leading order in the thickness of the brane \cite{Carter2}. These type of corrections have been studied lately 
in the case of a static, axisymmetric brane on the background of a spherically symmetric black hole in 
arbitrary number of dimensions \cite{CF,FG}. 

This, so called, brane - black hole (BBH) system, in the infinitely thin case, was first introduced by Frolov 
\cite{Frolov} as a toy model for the study of merger- and topology changing transitions between higher 
dimensional black type solutions (or {\it phases}) of the Einstein field equations \cite{Kol1,Kol2,Kol3}. 
The model also turned out to be very useful through the AdS/CFT correspondence for the study of phase 
transitions in certain strongly coupled gauge theories \cite{MMT1, MMT2}, while the higher dimensional
generalizations of the Bernstein conjecture \cite{B1,B2}, the study of the stability of brane - black 
hole systems \cite{B3}, and the question of possible micro black hole formations in high energy collisions 
(like of the LHC) \cite{flactan} are also important directions which are dealing with similar setup, and 
provide motivation for the study of the curvature corrected problem. 

Stiffness (or thickness) corrections to the BBH system has been studied first in \cite{FG}, with a 
perturbative approach near the critical solution of the thin system in the Rindler zone. It was found 
that when the brane has more than 2 spatial dimensions, supercritical solutions behave quite 
differently from subcritical ones, and no evidence for the existence of such solutions has been found.
A possible explanation of this behavior was that stiffness corrections to the brane action break the 
symmetry between the super- and subcritical solutions, and quantum-gravitational effects might cure 
the problem.

The thickness corrected BBH system has been further studied in \cite{CF} within a more general  
perturbative approach that also looked away from the Rindler zone. In accordance with \cite{FG}, 
it was found that no regular subcritical thick solutions exist if the brane has more than 2 spacelike 
dimensions. This result however obtained a simple explanation with observing the fact that perturbative 
approaches break down around the corresponding thin solutions at the axis of the system, because the zero 
order (thin) solutions are not smooth there. As a consequence, the topology change transition in the 
thick system can not be studied in the general case  within a perturbative approach, and one needs to 
find a new, exact solution of the problem for Minkowski type topologies.

In the present paper we provide an exact numerical solution of the thick-BBH problem with arbitrary 
number of brane and bulk dimensions for both Minkowski- and black hole topologies. This work 
is an organic continuation of our previous perturbative approach and we kindly refer 
the reader to \cite{CF} for those definitions, notation and results that might be missing here and 
would make the present paper completely self-contained. 
 
The plan of the paper is as follows. In Sec.~II we run trough the same quick overview of the infinitely 
thin case that we provided in \cite{CF}, and reintroduce the most important parts of the BBH setup 
to make the paper self-contained. In Sec.~III we write up the curvature corrected DNG action and introduce 
the correction parameters. In Sec.~IV we derive the Euler-Lagrange equation and the equation of motion for 
the thick BBH system. In Sec.~V the regularity conditions are discussed while in Sec.~VI the
far distance asymptotic solution is obtained. In Sec.~VII we provide the exact, numerical solution of the 
problem in the near horizon region for both topologies, and in Sec.~VIII a topology change transition
is analyzed by considering the energy properties of a quasi-static thick brane evolution in the case
of a 5-dimensional brane in a 6-dimensional bulk.
 
\section{The thin BBH system setup}

Let us overview, in this section, the important properties of the BBH system, introduced in \cite{Frolov}, 
that we intend to study in the presence of a small brane thickness in the following sections. We consider 
static brane configurations in the background of a static, spherically symmetric bulk black hole. The 
metric of an $N$-dimensional, spherically symmetric black hole spacetime is
\begin{equation}
ds^2=g_{ab}dx^adx^b=-fdt^2+f^{-1}dr^2+r^2d\Omega_{N-2}^2\ ,
\end{equation}
where $f=f(r)$ and $d\Omega_{N-2}^2$ is the metric of an $N-2$ dimensional unit sphere. One can define 
coordinates $\theta_i (i=1,\dots, N-2)$ on this sphere with the relation
\begin{equation}
d\Omega_{i+1}^2=d\theta_{i+1}^2+\sin^2\theta_{i+1} d\Omega_i^2 \ .
\end{equation}
The explicit form of $f$ is not important, it is only assumed that $f$ is zero at the horizon $r_0$, and it 
grows monotonically to $1$ at the spatial infinity $r\rightarrow \infty$, where it has the asymptotic form 
\cite{Tangherlini},
\begin{equation}\label{f}
f=1-\left(\frac{r_0}{r}\right)^{N-3}\ .
\end{equation}

In the zero thickness case, the test brane configurations, in an external gravitational field, can be obtained 
by solving the equation of motion coming from the Dirac-Nambu-Goto action \cite{Dirac, Nambu, Goto},
\begin{equation}\label{action0}
S=\int d^D\zeta\sqrt{-\mbox{det}\gamma_{\mu\nu}}\ , 
\end{equation}
where $\gamma_{\mu\nu}$ is the induced metric on the brane
\begin{equation}
\gamma_{\mu\nu} =g_{ab}\frac{\partial x^a}{\partial \zeta^{\mu}}
\frac{\partial x^b}{\partial \zeta^{\nu}}\ ,
\end{equation}
and $\zeta^{\mu}(\mu=0,\dots ,D-1)$ are coordinates on the brane world sheet. The brane tension does not enter 
into the brane equations, thus for simplicity it can be put equal to $1$. It is also assumed that the brane is 
static and spherically symmetric, and its surface is chosen to obey the equations
\begin{equation}
\theta_D=\dots =\theta_{N-2}=\pi/2\ .
\end{equation}
With the above symmetry properties the brane world sheet can be defined by the function $\theta_{D-1}=\theta(r)$ 
and we shall use coordinates $\zeta^{\mu}$ on the brane as
\begin{equation}
\zeta^{\mu}=\{t,r,\phi_1,\dots,\phi_{n}\}\quad \mbox{with} \quad n=D-2 \ . 
\end{equation}
With this parametrization the induced metric on the brane is
\begin{equation}
\gamma_{\mu\nu} d\zeta^{\mu}d\zeta^{\nu}=-fdt^2+\left[\frac{1}{f}+r^2{\dot\theta}^2\right]dr^2+r^2\sin^2\theta d\Omega_n^2,
\end{equation}
where, and throughout this paper, a dot denotes the derivative with respect to $r$, and the action (\ref{action0}) reduces 
to
\begin{eqnarray}
S&=&\Delta t \mathcal{A}_n\int\mathcal{L}_0\ dr\ ,\\
\mathcal{L}_0&=&r^n\sin^n\theta\sqrt{1+fr^2{\dot\theta}^2}\ ,\label{L0} 
\end{eqnarray}
where $\Delta t$ is the interval of time and $\mathcal{A}_n=2\pi^{n/2}/\Gamma(n/2)$ is the surface area of a unit $n$-dimensional 
sphere.

\section{The thick brane action}
In the case of a thick brane, the curvature corrected effective brane action is obtained by Carter and Gregory 
\cite{Carter2}, and  the corrections to the thin DNG action are induced by small thickness 
perturbations
\begin{equation}\label{action1}
S=\int d^D\zeta\sqrt{-\mbox{det}\gamma_{\mu\nu}}\left[-\tfrac{8\mu^2}
{3 \ell}(1+C_1R+C_2K^2)\right],
\end{equation}
where $R$ is the Ricci scalar, $K$ is the extrinsic curvature scalar of the brane and the coefficients $C_1$ and 
$C_2$ are expressed by the wall thickness parameter $\ell$ as
\begin{equation}
C_1=\frac{\pi^2-6}{24}\ell^2\ , \qquad C_2= -\frac{1}{3}\ell^2 .
\end{equation}
The parameter $\mu$ is related to the thickness by 
\begin{equation}
\ell=\frac{1}{\mu\sqrt{2\lambda}}
\end{equation}
which originates from a field theoretical domain-wall model, where $\mu$ is the mass parameter and $\lambda$ is the 
coupling constant of the scalar field.

After integrating out the spherical symmetric part and the time dependence on the introduced static, spherically 
symmetric, higher dimensional black hole background, one obtains 
(see also \cite{CF})
\begin{eqnarray}\label{S}
S&=&\Delta t \mathcal{A}_n\int\mathcal{L}\ dr\ ,\\
\mathcal{L}&=&-\frac{8\mu^2}{3\ell}\mathcal{L}_0[1+\ve\delta]\ ,\label{L} 
\end{eqnarray}
where we introduced the notations 
\begin{eqnarray}\label{delta}
\ve=\frac{\ell^2}{L^2},\qquad \delta&=&aK^2+bQ\ ,
\end{eqnarray}
with
\begin{equation}\label{ab}
Q= K^a_bK^b_a,\quad a=\frac{\pi^2-14}{24}L^2,\quad b=\frac{6-\pi^2}{24}L^2\ ,
\end{equation}
and $L$ is the relevant dynamical length scale of the system which has to be large compared to 
the thickness parameter $\ell$. The explicit expressions of the curvature scalars $K$ and $Q$ are 
given in (35) and (36) of \cite{CF}.

\section{The Euler-Lagrange equation}
The curvature corrected DNG-brane action has a dependence on the second derivative of $\theta$, thus the 
Euler-Lagrange equation of the problem reads
\begin{equation}\label{mel}
\frac{d^2}{dr^2}\left(\frac{\partial \mathcal{L}}{\partial \ddot\theta}\right)
-\frac{d}{dr}\left(\frac{\partial \mathcal{L}}{\partial \dot\theta}\right)
+\frac{\partial \mathcal{L}}{\partial \theta}=0\ .
\end{equation}
Plugging the thickness corrected Lagrangian density (\ref{L}) into (\ref{mel}) we get
\begin{eqnarray}\label{lagp}
\!\!\!\!\!\!\!\!\!\!\!\!\!\!\!\!\!\!\!\!
0&=&\frac{d}{dr}\left(\frac{\partial \mathcal{L}_0}{\partial \dot\theta}\right)
-\frac{\partial \mathcal{L}_0}{\partial \theta}\nonumber\\
\!\!\!\!\!\!\!\!\!
&-&\ve\left[\frac{d^2}{dr^2}\left(\frac{\partial (\mathcal{L}_0\delta)}{\partial \ddot\theta}\right)
-\frac{d}{dr}\left(\frac{\partial (\mathcal{L}_0\delta)}{\partial \dot\theta}\right)
+\frac{\partial (\mathcal{L}_0\delta)}{\partial \theta}\right],
\end{eqnarray}
from which the actual equation of motion becomes
\begin{eqnarray}\label{fulleq}
\theta^{(4)}&+&T_1\theta^{(3)}
+T_2(\ddot\theta,\dot\theta,\theta,f^{(3)},\ddot f,\dot f,f,r)=0\ ,
\end{eqnarray}
where
\begin{eqnarray}\label{T1}
T_1&=&\frac{1}{rfF^2}\left[\right.
(2n+4)f+4r\dot f+2nrf\cot\theta\dot\theta\nonumber\\
&+& r^2f\left[2(n-3)f-r\dot f\right]\dot\theta^2\nonumber\\
&+&2nr^3f^2\cot\theta\dot\theta^3-10r^3f^2\ddot\theta
\left. \right]\ ,
\end{eqnarray}
and $T_2$ is given in the appendix.

The goal of this paper is to provide a regular, exact, numerical solution of (\ref{fulleq}) in 
arbitrary brane and bulk dimensions for both Minkowski- and black hole type topologies, and 
study the properties of a quasi-static brane transition between the two type of solutions. 

\section{Regularity conditions}

\subsection{Minkowski embedding case}
As a first step we consider the asymptotic behavior of (\ref{fulleq}) near the axis of the system 
($\theta=0$) in the Minkowski embedding case, where the perturbative approaches failed to provide 
regular solutions in the dimensions $n\geq 2$. The asymptotic equation (obtained by taking the series 
expansion of (\ref{fulleq}) around $\theta=0$) reads
\begin{equation}\label{theta0eq}
\frac{s_3}{\theta^3}+\frac{s_2}{\theta^2}+\frac{s_1}{\theta}+s_0+\dots =0\ ,
\end{equation}
where the functions $s_1$, $s_2$, and $s_3$ are given in the appendix.
For a regular solution one needs to require $s_3$, $s_2$ and $s_1$  
to disappear at $r_1$. 

The requirement $s_3|_{r_1}=0$ is automatically satisfied in the cases of $n=0$ and $n=2$.
For other dimensions we obtain the formula 
\begin{equation}\label{regt1}
\dot\theta|_{r_1}=\pm\left.\sqrt{\frac{-(b+an)}{2(2a+b-an)r^2f}}\right|_{r_1}\ .
\end{equation}
We are looking for real solutions, hence we have the condition 
\begin{equation}\label{cond}
\frac{-(b+an)}{2(2a+b-an)}\geq 0\ .
\end{equation}
Plugging the explicit values of $a$ and $b$ into (\ref{cond}) one can find that 
the inequality is satisfied for all $n\geq 3$. In the case of $n=1$, the regularity 
condition (\ref{regt1}) has no real solution for $\dot\theta|_{r_1}$, which is a rather
unexpected result. Note that this result is independent of the explicit values 
of $a$ and $b$.

From the requirements $s_2|_{r_1}=0$ and $s_1|_{r_1}=0$ we obtain 
unique conditions for $\ddot\theta|_{r_1}=g(\dot\theta|_{r_1})$ 
and $\theta^{(3)}|_{r_1}=h(\dot\theta|_{r_1},\ddot\theta|_{r_1})$, 
where the functions $g$ and $h$ are given in the appendix. 

Accordingly, one can always find a regular solution for 
the equation (\ref{fulleq}) at the axes ($\theta|_{r_1}=0$) 
if $n\geq 3$. In this case the initial conditions are also uniquely fixed 
by the regularity conditions. In the cases of $n=0$ and 
$n=2$, regular solutions do exist, however one has a freedom in fixing 
the initial condition $\dot\theta|_{r_1}$. Having fixed $\dot\theta|_{r_1}$,
the remaining conditions are uniquely determined. 

In the $n=1$ case, the regularity requirements can not be satisfied with the 
presented method. It is interesting however, that in this case the perturbative 
approach worked well, and provided unique, regular solution to the problem. 
Furthermore, as explicitly constructed field theoretical domain-wall models 
\cite{Fl1,Fl2} clearly show the existence of regular solutions in this case, 
we believe that the obtained irregularity has no any physical meaning, rather 
it originates from some kind of break down of the applied method in the special 
case of $n=1$. A similar problem occurred for example during the integration 
of the far distance solution in \cite{CF}, where in the $n=3$ case, the solution
developed a resonant term. Unfortunately, since the equation of motion in the 
present exact case is so complicated and so highly non-linear, it is very difficult 
to follow up analytically the source of this irregular behavior, and thus 
its origin is presently not clear to the author.

\subsection{Black hole embedding case}

In the black hole embedding case the perturbative approach provided 
regular thick solutions in every brane dimensions. Nevertheless, 
for a complete analysis, we examine the asymptotic behavior of 
(\ref{fulleq}) near the black hole horizon and compute the exact, 
numerical solution of the problem.

As $r\rightarrow r_0$, (i.e.~on the horizon), the metric function $f(r)$ 
goes to zero (see (\ref{f})) but all of its derivatives are nonzero
and finite. We can thus consider the asymptotic behavior of (\ref{fulleq}) 
by taking its series expansion around $f(r)=0$, and get
\begin{equation}\label{f0eq}
\frac{y_2}{f^2}+\frac{y_1}{f}+y_0+\dots =0\ ,
\end{equation}
where $y_1$ and $y_2$ are given in the appendix. Similarly to the 
Minkowski embedding case, from $y_2$ one can obtain a unique condition 
for $\ddot\theta|_{r_0}=G(\dot\theta|_{r_0})$, and another for 
$\theta^{(3)}|_{r_0}=H(\dot\theta|_{r_0},\ddot\theta|_{r_0})$ from $y_1$. 
The functions $G$ and $H$ are given in the appendix. In addition 
one can fix $\theta|_{r_0}$ arbitrarily between $0$ and $\pi/2$ (which 
we will actually do later to consider a quasi-static brane 
evolution), however one still remains free to fix the initial condition  
$\dot\theta|_{r_0}$. 

Accordingly, the black hole embedding problem can  be regularly solved in 
any dimension, but the regularity requirements do not fix the initial 
conditions uniquely. For being able to provide an exact, numerical solution, 
we will adopt the initial condition for $\dot\theta|_{r_0}$ from the perturbative 
approach \cite{CF}, where it was uniquely fixed by regularity considerations. This 
choice is supported by the fact that in the black hole embedding case 
there is no problem with the perturbative approach around the thin solution, 
and the exact thick solution is dominated by the linear term in the 
$\ve$-expansion on the horizon. The contributions of the higher order terms to the 
$\dot\theta|_{r_0}$ initial condition are negligible.

\section{The far distance solution}

As $r\rightarrow\infty$ the far distance solution can be obtained from the 
condition that the brane behaves asymptotically as a $D-1$-dimensional plane. 
We look for the solution in the form
\begin{equation}
\theta=\frac{\pi}{2}+\nu(r),
\end{equation}
where $\nu(r)$ is small and we keep only its linear terms in (\ref{fulleq}). 
The asymptotic equation reads
\begin{equation}\label{rinfeq}
\nu^{(4)}+\tfrac{2(n+2)}{r}\nu^{(3)}-k\ddot\nu-\tfrac{k(n+2)}{r}\dot\nu-\tfrac{nk}{r^2}\nu=0\ ,
\end{equation}
where $k=(8(a+b)\ve)^{-1}$.

It is instructive to compare (\ref{rinfeq}) to the asymptotic form of the 
perturbation equation (57) of \cite{CF}. As one would expect, the terms 
up to the second derivative are identical, while the $3rd$- and $4th$-order 
terms in (\ref{rinfeq}) are the explicit correspondents of the source term 
that appears in (57) of \cite{CF}. 

For the solution, we obviously expect the same behavior that 
we obtained in the perturbative approach, although (\ref{rinfeq}) can not 
be integrated in a simple closed form. In order to check our expectation, we map 
analytically the point at infinity into the origin using the inversion 
transformation method (see eg. \cite{BO})
\begin{eqnarray}
r&=&\frac{1}{t}\ ,\nonumber\\
\frac{d}{dr}&=&-t^2\frac{d}{dt}\ ,\nonumber\\
\frac{d^2}{dr^2}&=&t^4\frac{d^2}{dt^2}+2t^3\frac{d}{dt}\ ,\nonumber
\end{eqnarray}
and so on, and look for the solution of the transformed asymptotic equation
\begin{equation}\label{t0eq}
\nu^{(4)}+\tfrac{2(4-n)}{t}\nu^{(3)}-\tfrac{k}{t^4}\nu''+\tfrac{nk}{t^5}\nu'-\tfrac{nk}{t^6}\nu=0
\end{equation}
as $t\rightarrow 0$. In (\ref{t0eq}) a prime denotes a derivative with 
respect to $t$.

Even though we observe that (\ref{t0eq}) has an irregular singular point 
at $t=0$, one can easily find a completely regular, exact solution 
\begin{equation}
\nu(t)=Pt \quad \Longrightarrow\quad \nu(r)=\frac{P}{r}+\dots\ .
\end{equation}
This is exactly the far distance behavior that we expected, 
and the coefficient $P$, just as in the thin case, can be referred as the 
distance of the brane from the equatorial plane at infinity.

\section{Numerical solution in the near horizon region}

After obtaining the regularity conditions and providing the far distance
solution of the exact curvature corrected problem, we consider the numerical 
solution of (\ref{fulleq}) in the near horizon region. Since we found that 
the initial conditions are not completely fixed for Minkowski embedding 
topologies in the cases of $n=0$ and $2$, and also concluded that one can not 
satisfy the regularity condition (\ref{cond}) in the $n=1$ (sheet) case with 
the presented method, we choose to illustrate the numerical solution in the 
case of a $D=5$-dimensional ($n=3$) thick brane embedded in a 6-dimensional 
bulk. 

Within the perturbative description (see \cite{CF} for details) we could 
not approach the curvature singularity arbitrary close unless we adjusted 
the thickness of the brane accordingly by changing the value of the 
perturbation parameter $\ve$ in every step. Instead, the method we followed was 
fixing $\ve$ to its maximum value, which is equivalent to choose the thickest 
possible brane configuration for a previously obtained length scale parameter 
$L$ that had been determined by the boundary position ($\theta_0$ or $r_1$) 
of the brane. As we approached the singularity, the absolute maximum of the 
perturbations were growing and finally we arrived to a minimal $\theta_0$ 
value, at which the applied perturbation method reached its limitation.

With the present, exact solution however, nothing prevents us to approach the 
singularity as close as we please, and the only restriction we have to bear 
in mind is the validity of the curvature corrected DNG-brane action 
(\ref{action1}), i.e.
\[
\frac{\ell}{L}\ll1\ .
\]
By keeping the concept of obtaining the thickest possible brane configuration 
for given boundary data, just as in the perturbative approach, we choose to 
fix the perturbation parameter $\ve$ to its maximum value obtained from the 
condition
\[
\left.\frac{\ell}{L}\right|_{max}\sim 0.1\quad\Longrightarrow
\quad \ve_{max}\equiv \left.\frac{\ell^2}{L^2}\right|_{max}\simeq 0.01\ .
\]
This requirement provides us the thickest possible brane configurations for 
every given length scale determined by the boundary data $\theta_0$ with no 
restriction on how close we are to the curvature singularity.

\subsection{Minkowski embedding solutions}

In the perturbative approach, regular thick brane solutions did not exist for Minkowski topologies
if the brane had more than two spacelike dimensions, i.e. for $n\ge 2$. In Fig.~1, we have plotted 
a set of exact thick brane solutions with varying boundary condition ($\theta_0$ for black hole embeddings
and $r_1$ for Minkowski embeddings) in the case of a 5-dimensional ($n=3$) brane embedded in a 6-dimensional
bulk.
\begin{figure}[!ht]
\noindent\hfil\includegraphics[scale=1]{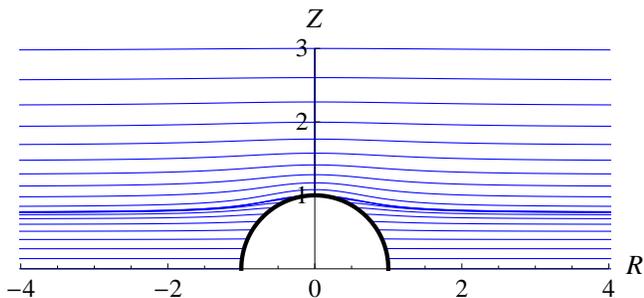} 
\caption{The picture shows a sequence of subcritical and supercritical thick brane solutions 
in the case when a $D=5$ dimensional brane embedded into a $N=6$ dimensional bulk. The different 
configurations belong to different initial conditions of $r_1$ and $\theta_0$, respectively. 
For simplicity, the bulk black hole's horizon radius is put to be 1, and $R$ and $Z$ are the standard 
cylindrical coordinates.}
\end{figure}

There is no apparent difference on Fig.~1 if we compare it to the corresponding thin solutions. This is 
simply because we are considering second order curvature corrections to the thin brane action so the effects
are indeed very tiny. To make these effects visible we plot a sequence of the difference function 
\[
 \Delta\theta(r)=\theta(r)-\theta_{DNG}(r) ,
\]
of the corresponding thick and thin exact solutions, which is the analog of the perturbation function 
$\varphi(r)$ defined in (41) of \cite{CF}.  

The qualitative behavior of the individual $\Delta\theta$ curves are very similar to the $\varphi$ 
perturbations that we obtained in the $n=1$ case in \cite{CF}. The curves that are very close to the 
horizon has a sharp negative maximum and change sign before decaying. On the actual shape of the curves 
however we can observe the effect of the higher orders as having an extra oscillatory pattern. 
In Fig.~2 we have plotted a $\Delta\theta$ curve corresponding to $r_1=1.001$ boundary condition. 
\begin{figure}[!ht]
\noindent\hfil\includegraphics[scale=1]{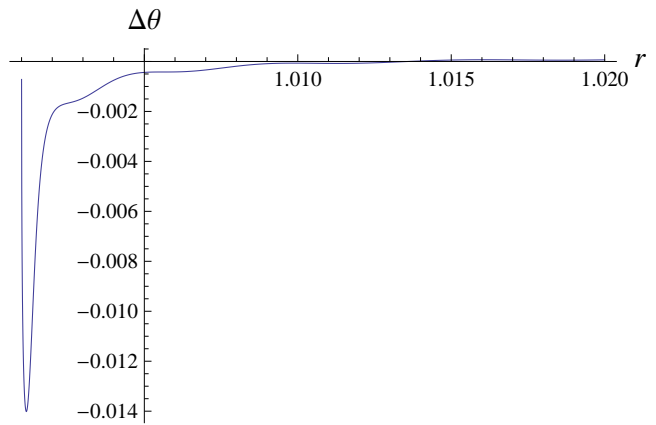} 
\caption{The picture shows a near horizon $\Delta\theta$ curve with minimum horizon distance $r_1=1.001$.}
\end{figure}

If we go away from the horizon, the difference function becomes more spread and doesn't change sign
before decaying. The negative maximum and the oscillatory extra pattern are still present. In Fig.~3 
we have plotted a $\Delta\theta$ curve corresponding to $r_1=2$ boundary condition. 

\begin{figure}[!ht]
\noindent\hfil\includegraphics[scale=1]{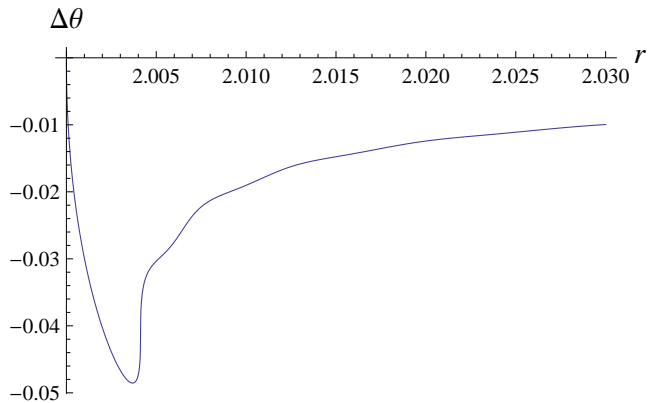} 
\caption{The picture shows a $\Delta\theta$ curve a bit further away from the black hole horizon with 
minimum horizon distance $r_1=2$.}
\end{figure}

As we can see, there is no essential difference in the individual $\Delta\theta$ curves  
compared to the perturbation function $\varphi$ obtained in \cite{CF} in the $n=1$ case. A remarkable 
difference appears however, if we take a look on the absolute maximum of the curves as we increase the 
minimum distance parameter $r_1$. In Fig.~4 we have plotted a sequence of $\Delta\theta$ curves.
\begin{figure}[!ht]
\noindent\hfil\includegraphics[scale=1]{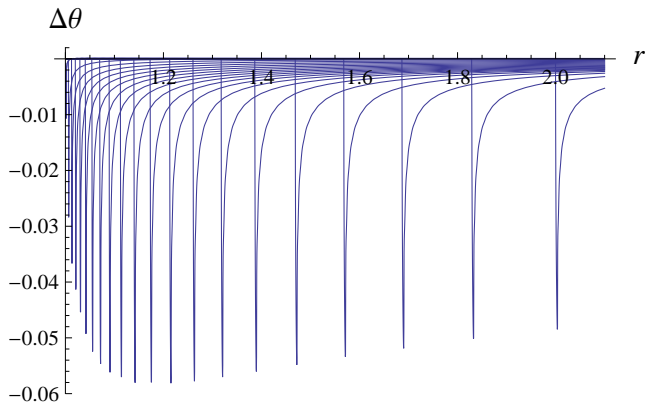} 
\caption{The picture shows a sequence of $\Delta\theta$ curves with increasing minimum distance parameter 
in the $1.001\le r_1\le 2$ interval.}
\end{figure}
It can be seen that the absolute maximum of the difference function curves does not decay monotonically with
increasing distance from the horizon. Instead, it has a growing tendency until it reaches a maximum around 
$r_1\simeq 1.15$, after which it starts decaying monotonically as one would expect. The concrete numerical
values of the absolute maximum of the curves are not relevant as they depend on the chosen normalization 
condition, i.e.~on the maximum length parameter $L$. In analyzing the Minkowski embedding solutions, we have 
used the normalization parameter that belongs to the boundary condition $\theta_0=\tfrac{\pi}{18}$. 
 
Apart, of course, from the very existence of the thick solutions in the $n\geq 2$ cases, the above 
non-monotonic decaying is an interesting and clear difference between the exact and perturbative 
solutions.

\subsection{Black hole embedding solutions}
The non-monotonicity in the maximum values of the difference function $\Delta\theta$ with respect to the 
initial value parameter $\theta_0\in(0,\tfrac{\pi}{2}]$ is also present in the 
black hole embedding case. As we mentioned earlier, the regularity requirements do not fix the 
$\dot\theta|_{r_0}$ conditions for the black hole case, and to provide a set of exact solutions 
we took the missing conditions from our previously obtained perturbative results \cite{CF}.

We also discussed that with the presented exact solution we can approach the curvature singularity in 
principle as close as we please, and it is only the capacity of our computing facility that can put some 
limitation on us. Taking into account our limits, we have chosen to solve our equation starting from the
parameter $\theta_0=\tfrac{\pi}{900}$, which is a significant increase compared to the limit
$\tfrac{\pi}{18}$, what we had within the perturbative approach. The corresponding $\Delta\theta$ curve is 
plotted on Fig.~5.
\begin{figure}[!ht]
\noindent\hfil\includegraphics[scale=1]{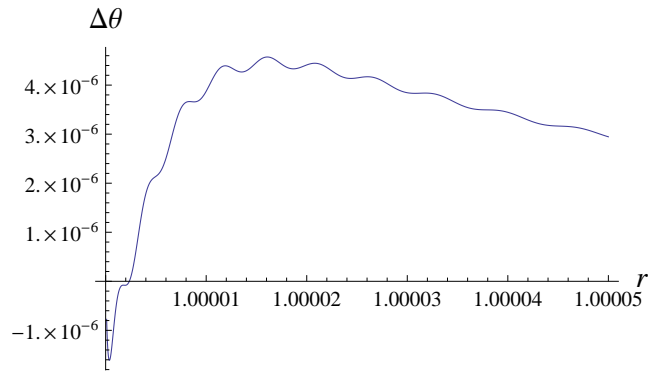} 
\caption{The $\Delta\theta$ function in the black hole embedding case with $\theta_0=\tfrac{\pi}{900}$
at the very near horizon region.}
\end{figure}

The oscillatory pattern of the higher order effects are clearly present just as in the Minkowski
embedding case, and before decaying eventually, the curve change its sign 3 times. This can not be seen 
on Fig.~5 as the amplitude is decaying very rapidly and the second and third sign change takes place further
away from the horizon.  

As we increase the $\theta_0$ parameter from $\tfrac{\pi}{900}$ to $\tfrac{\pi}{2}$ (i.e. changing the 
initial position of the brane on the horizon from the near pole region to the equator) the shapes and the 
maximum of the corresponding $\Delta\theta$ curves are changing. The tendency in these features is not 
so simple as in the Minkowski embedding case, for example there are several local extrema of the maximum
with respect to $\theta_0$. Nevertheless, to illustrate some of the solutions, we picked out three examples 
with initial parameters $\theta_0\sim\tfrac{\pi}{6}$, $\tfrac{\pi}{3}$ and $\tfrac{\pi}{2}$. The 
corresponding curves are plotted on Figs.~6, 7 and 8, respectively.
\begin{figure}[!ht]
\noindent\hfil\includegraphics[scale=1]{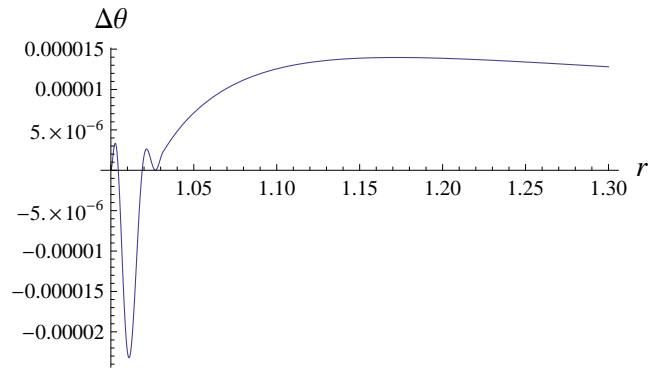} 
\caption{The $\Delta\theta$ function in the black hole embedding case with $\theta_0\simeq\tfrac{\pi}{6.6}$.}
\end{figure}
\begin{figure}[!ht]
\noindent\hfil\includegraphics[scale=1]{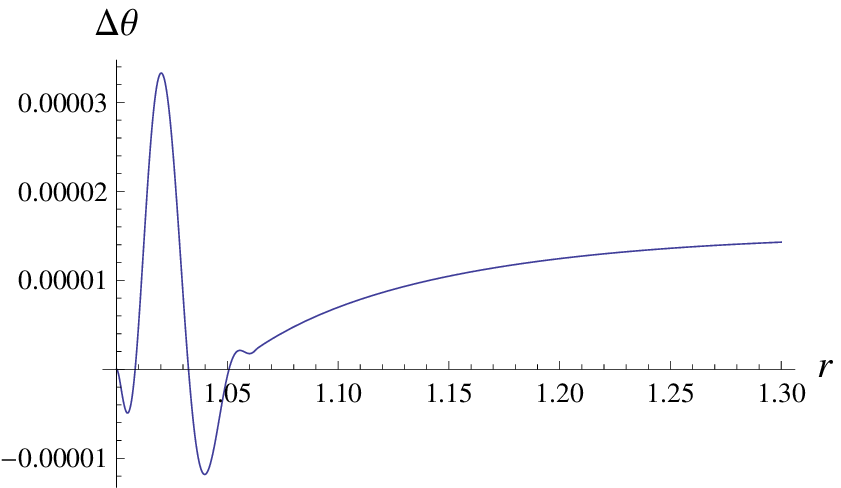} 
\caption{The $\Delta\theta$ function in the black hole embedding case with $\theta_0\simeq\tfrac{\pi}{2.85}$.}
\end{figure}
\begin{figure}[!ht]
\noindent\hfil\includegraphics[scale=1]{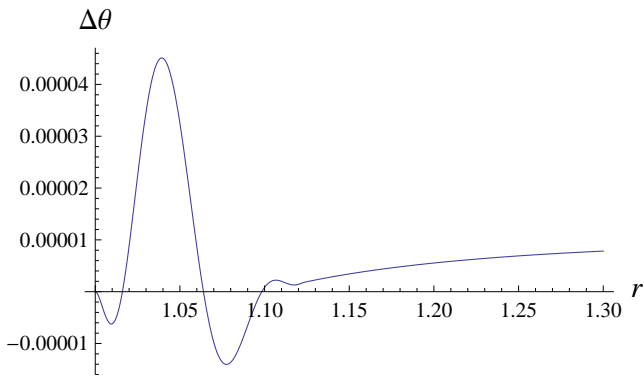} 
\caption{The $\Delta\theta$ function in the black hole embedding case with $\theta_0\simeq\tfrac{\pi}{2.2}$.}
\end{figure}

For all curves in the black hole embedding case, the normalization has been calculated according to the 
length scale parameter corresponds to $\theta_0=\tfrac{\pi}{900}$.

\section{Quasi-static phase transition}

Having in hand the exact numerical solution of the curvature corrected BBH problem, we are in the
position to analyze the topology changing transition between the black hole and Minkowski embedding 
{\it phases} by considering the energy properties of a quasi-static brane evolution from an 
initially equatorial configuration. This method has been introduced first by Flachi {\it et al.} in 
\cite{Flachi}, and its details for our purposes have been explained in \cite{CF}. 

The idea is very simple, one compares the energy of a thick brane configuration that belongs to
a chosen initial parameter $\theta_0$ (in the black hole embedding case) or $r_1$ (in the Minkowski 
embedding case) to the energy of the equatorial configuration, i.e.~to the brane with 
$\theta_0=\tfrac{\pi}{2}$.  If one plots the energy difference $\Delta E$ introduced this way as a 
function of the cylindrical distance parameter defined as
\[
Z_{\infty}=r_{\infty}\cos\theta(r_{\infty}) 
\]
(where $r_{\infty}$ denotes a certain cut off distance very far from the horizon) for every initial 
parameter $\theta_0$ and $r_1$, then the obtained plot exhibits a loop, or instability zone, that is 
a typical sign of a first order phase transition in dynamical systems. For further details please refer 
to \cite{CF} or \cite{Flachi}.

In Fig.~9 we have plotted the $\Delta E(Z_{\infty})$ curves for the presented 5-dimensional brane solutions 
in a 6-dimensional bulk. The red curve represents the evolution of the brane in the black hole embedding 
phase, while the blue belongs to the evolution in the Minkowski embedding phase. It can be seen on the 
picture that something interesting happens around $Z_{\infty}\simeq 0.8$, where the two curves seems to touch 
each other.
\begin{figure}[!ht]
\noindent\hfil\includegraphics[scale=1]{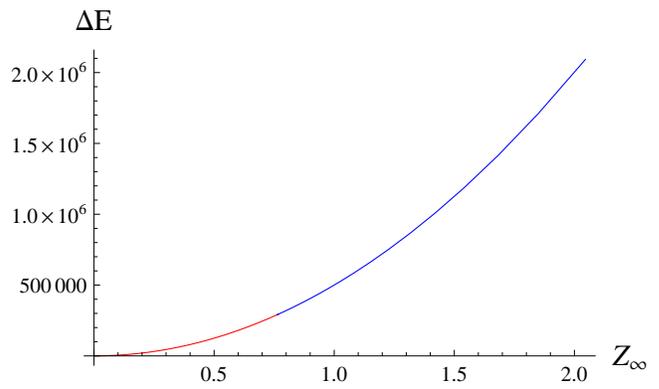} 
\caption{The picture shows the $\Delta E(Z_{\infty})$ curves of a quasi-static thick brane evolution
in the case of a 5-dimensional brane in a 6-dimensional bulk. The red curve belongs to the black hole
embedding-, while the blue curve to the Minkowski embedding evolution.}
\end{figure}

If we enlarge this part on Fig.~10, it becomes apparent that indeed the two curves overlap in the
$0.76\lesssim Z_{\infty}\lesssim 0.77$ interval, which clearly shows that there is an energy degeneracy 
in the brane evolution at the near singularity region. In fact this overlap exhibits a real one-dimensional
loop in the sense that both curves has a turn-back point in this region. The black hole embedding (red) curve
increases from zero until it reaches its maximum around $Z_{\infty}\simeq 0.77$, where it turns back and start 
decreasing and converges to some point in the overlap region. The Minkowski embedding (blue) curve starts 
decreasing from the same point (this point represents the curvature singularity) until $Z_{\infty}\simeq 0.76$, where
it turns back and starts increasing to infinity. Unfortunately, since being a one-dimensional loop, it can 
not be seen on the graph.
\begin{figure}[!ht]
\noindent\hfil\includegraphics[scale=1]{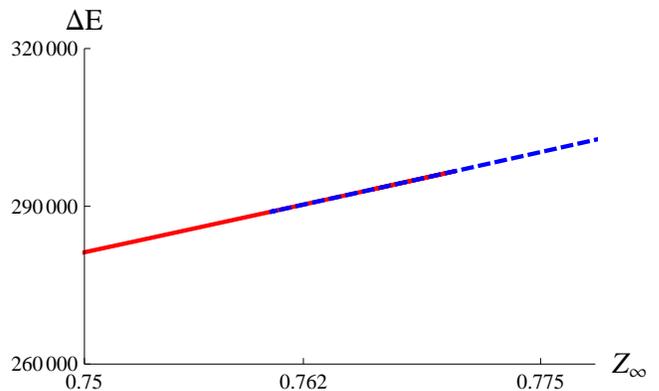} 
\caption{The enlargement of the overlap region on Fig.~9.}
\end{figure}

In any case, the above property of the brane evolution demonstrates that the corresponding topology changing
transition remains a first order one in the curvature corrected BBH system, just as it was in the infinitely 
thin case. 

\section{Conclusions}
In the present paper we further studied the effects of higher order, curvature
corrections to the dynamics of higher dimensional brane black hole systems. Since earlier
results \cite{CF,FG} clearly showed that perturbative approaches fail to provide
regular solutions near the axis of the system in Minkowski type topologies, we 
considered a different, exact, numerical approach to the problem. We analyzed 
the asymptotic properties of the complete $4th$-order equation of motion of the 
thick BBH system, provided its asymptotic solution for far distances, and obtained
regularity conditions in the near horizon region for both Minkowski and black hole
embeddings.

We showed that the requirement of regularity for the thick solution defines 
almost completely the boundary conditions for the Euler-Lagrange 
equation in the Minkowski embedding case. The only exceptions are the brane 
configurations with 1, 2 and 3 spacelike dimensions. In the case of 1 and 3
spacelike dimensions regular solutions of the problem exist, however, which 
is an unexpected result, with 2 spaclike dimensions the problem can not be
solved with the applied method. Nevertheless it is obvious that the lack of a 
regular solution in this case should not have any physical reason as a regular 
perturbative solution clearly exist in accordance with 
explicitly constructed field theoretical domain wall solutions \cite{Fl1,Fl2}
to the problem.

For the black hole embedding solutions, the boundary conditions are not completely 
determined form the regularity requirements, nevertheless, since the perturbative 
approach, presented in \cite{CF}, is valid for this case, the missing conditions can 
be borrowed from those results. 

Having in hand all the boundary conditions we provided a set of exact, numerical solutions
in the near horizon region for both Minkowski and black hole topologies in the case when 
a 5-dimensional brane embedded into a 6-dimensional bulk. Based on the obtained solutions 
we discussed the properties of a topology changing transition in the system, and concluded 
that the transition remains first order just as it was in the thin model.

The main results of the paper are the construction of the exact numerical solution for the 
thick BBH problem in any dimensions (apart from the $n=1$ special case), and the clarification 
of the question whether the simplest, higher order, curvature corrections can modify the order 
of the phase transition in the system.

The obtained results are relevant in several research areas that are considering similar setup 
to the BBH system. A few of them were mentioned in the introduction section. 

\acknowledgments
I am grateful for valuable discussions with Antonino Flachi. Most part of the calculations were 
performed and checked using the computer algebra program MATHEMATICA 7. The research was supported 
by the Hungarian National Research Fund, OTKA No.~K67790 grant.

\appendix*
\section{Coefficient Functions}
\begin{widetext}
{\small
\begin{eqnarray}\label{T2}
T_2&=&\frac{1}{64(a+b)\varepsilon f^2r^4F^4}\left[\right.-8 \dot\theta^9 \varepsilon f^6 (a + b + a n) (-1 
+ n^2) r^9 \\
&-& 2 \dot\theta^2 f^4 r^5 (240 (a + b) \ddot\theta^2 \dot\theta \varepsilon (-3 + n) r^2 - 
480 (a + b) \ddot\theta^3 \varepsilon r^3 + 8 \ddot\theta \dot\theta^2 \varepsilon r (-71 (a + b) \nonumber\\
&+& 45 a n + 42 b n + 2 (a - 2 b) n^2 + 24 (a + b) \ddot\theta n r^2 \cot\theta) 
+ 4 \dot\theta^3 \varepsilon (b (-80 + (69 - 7 n) n) \nonumber\\
&+& a (-80 + n (54 + (29 - 3 n) n)) + 2 \ddot\theta (a (37 - 17 n) + b (25 - 8 n)) n r^2 \cot\theta)
\nonumber\\
&+&  4 \dot\theta^6 n r^3 \cot\theta (2 \varepsilon n (b - a n) - r^2 + 
\varepsilon r (-3 \dot f (a + 2 b + 3 a n) - 4 a \ddot f r) 
\nonumber\\
&-& 2 \varepsilon (b - a (-2 + n)) (-2 + n) \csc\theta^2) 
+ 4 \dot\theta^5 r^2 (\varepsilon n (-a (-25 + n) n + b (-3 + 11 n)) 
\nonumber\\
&+& (5 + 4 n) r^2 + \varepsilon r (\dot f (2 b (8 + n) + a (13 + n (27 + 5 n))) + 
r (\ddot f (5 a - 3 b - 2 a n) - 2 a f^{(3)} r)) 
\nonumber\\
&+& \varepsilon n (b (13 - 11 n) + a (28 + (-27 + n) n)) \csc\theta^2) + 
 4 \dot\theta^4 r (\varepsilon n (b (57 - 31 n) 
\nonumber\\
&+& a (89 - n (26 + 19 n))) \cot\theta +\ddot\theta r^2 (2 \varepsilon n (5 a n + b (-1 + 4 n)) + r^2 
+ \varepsilon r (3 \dot f (a + 2 b + 3 a n) 
\nonumber\\
&+& 4 a \ddot f r) + 2 \varepsilon (a (6 - 5 n) + b (5 - 4 n)) n \csc\theta^2))
+ \dot f \dot\theta^7 r^5 ((3 a + b) \dot f \varepsilon (1 + n) r + 2 (2 a \varepsilon n^2 
\nonumber\\
&+&  r^2 + (a + b) \ddot f \varepsilon r^2 - 2 a \varepsilon (-1 + n) n \csc\theta^2))) 
+  2 f^2 r (3 (2 a + b) \dot f^3 \dot\theta^7 \varepsilon r^9 
\nonumber\\
&-&  23 (a + b) \dot f^2 \dot\theta^6 \varepsilon n r^7 \cot\theta + 
 8 \ddot\theta \varepsilon r (a (2 + n) (4 + 5 n) + b (8 + n (15 + 4 n)) 
\nonumber\\
&+&  6 (a + b) \ddot\theta n r^2 \cot\theta) + 8 \dot\theta \varepsilon (n^2 (3 b + a (2 + n)) 
- 90 (a + b) \ddot\theta^2 \dot f r^5 + \ddot\theta n (b (25 + 8 n) 
\nonumber\\
&+& a (28 + 17 n)) r^2 \cot\theta) - 2 \dot f \dot\theta^5 r^5 (2 (a + 4 b) \varepsilon n^2 
+ 9 r^2 + 3 \varepsilon r (\dot f (5 b (-1 + n) + a (4 + 11 n)) 
\nonumber\\
&+& 8 (a + b) \ddot f r) -  2 (a + 4 b) \varepsilon (-1 + n) n \csc\theta^2) 
+ 4 \dot\theta^3 r^2 (\varepsilon n (b (3 - 9 n) + a n (-21 + 19 n)) 
\nonumber\\
&+& (-7 - 4 n) r^2 + \varepsilon r (2 r (\ddot f (b (9 + 4 n) + 2 a (5 + 7 n)) 
+ (5 a + 2 b) f^{(3)} r) + \dot f (2 (b (-78 
\nonumber\\
&+&  n (13 + 2 n)) + 2 a (-42 + n (-5 + 6 n))) + (47 a + 26 b) \ddot\theta n r^2 \cot\theta)) 
- \varepsilon (-1 + n) n (-9 b 
\nonumber\\
&+& a (-24 + 19 n)) \csc\theta^2) + 2 \dot\theta^4 r^3 (3 (7 a + 10 b) \ddot\theta \dot f^2 \varepsilon r^4 
+ 2 \varepsilon n r (\dot f (b (47 + 8 n) + a (74 + 53 n)) 
\nonumber\\
&+&  4 (5 a + 2 b) \ddot f r) \cot\theta + 6 n \cot\theta (\varepsilon n (-3 b + a n) 
+ 2 r^2 + \varepsilon (3 b - a (-4 + n)) (-2 + n) \csc\theta^2)) 
\nonumber\\
&+&  4 \dot\theta^2 r (\varepsilon n (21 b n + a (-6 + n (50 + n))) \cot\theta + 
 \ddot\theta r^2 (4 (a - 2 b) \varepsilon n^2 - 3 r^2 + 6 \varepsilon r (\dot f (a (-48 + n) 
\nonumber\\
&+& b (-46 + 3 n)) + (a + b) \ddot f r) - 4 (a - 2 b) \varepsilon (-1 + n) n \csc\theta^2))) - 
  4 \dot\theta^6 f^5 r^7 (2 \dot\theta \varepsilon (a + b 
\nonumber\\
&+&  a n) (-24 + n (17 + n)) + 2 \ddot\theta \varepsilon (-13 + 7 n) (a + b + a n) r 
- 2 \dot\theta^2 \varepsilon n (-13 + 7 n) (a + b 
\nonumber\\
&+& a n) r \cot\theta +  \dot\theta^3 r^2 (\varepsilon (1 + n) r (\dot f (3 (a + b) + 5 a n) 
+ 2 a \ddot f r) + 2 (2 \varepsilon n^2 (a + b + a n) 
\nonumber\\
&+&(1 + n) r^2 - 2 \varepsilon (-1 + n) n (a + b + a n) \csc\theta^2))) 
-  2 f (-9 (2 a + 3 b) \dot f^3 \dot\theta^5 \varepsilon r^8 
\nonumber\\
&-& 8 \varepsilon (b - a (-2 + n)) (-1 + n) n \cot\theta 
+  2 (-17 a + 4 b) \dot f^2 \dot\theta^4 \varepsilon n r^6 \cot\theta 
- 4 \ddot\theta r^2 (-r^2 
\nonumber\\
&+& 2 \varepsilon r (\dot f (20 a (2 + n) + b (40 + 17 n)) + (10 a + 9 b) \ddot f r) 
+ \varepsilon n (b - 7 a n) \cot\theta^2 + 8 a \varepsilon n \csc\theta^2) 
\nonumber\\
&-& 4 \dot\theta r (\varepsilon n^2 (-b + 7 a n) + (-2 - n) r^2 
+ 2 \varepsilon r (r (2 \ddot f (3 a (2 + n) + b (5 + n)) + (2 a + b) f^{(3)} r) 
\nonumber\\
&+& \dot f (2 a (2 + n) (2 + 3 n) +  b (8 + n (15 + n)) + (28 a + 19 b) \ddot\theta n r^2 \cot\theta)) 
+ \varepsilon (b 
\nonumber\\
&+&  a (8 - 7 n)) n^2 \csc\theta^2) + 2 \dot f \dot\theta^3 r^4 (\varepsilon n (-21 a n + b (3 + 4 n)) 
+ 7 r^2 + \varepsilon r (\dot f (2 a (62 + n) 
\nonumber\\
&+&  b (102 + 7 n)) + 12 b \ddot f r)+ \varepsilon n (b - 4 b n + 3 a (-8 + 7 n)) \csc\theta^2) 
-  2 \dot\theta^2 r^2 (-3 (46 a + 43 b) \ddot\theta \dot f^2 \varepsilon r^4 
\nonumber\\
&+& 2 n \cot\theta (-\varepsilon n (5 b + a n) + 4 r^2 + \varepsilon r (\dot f (74 a + 41 b + 4 (4 a + b) n) 
+ (5 a + 4 b) \ddot f r) 
\nonumber\\
&+& \varepsilon (-2 + n) (5 b + a (4 + n)) \csc\theta^2))) + 
  f^3 r^3 (-480 (a + b) \ddot\theta^2 \dot\theta \varepsilon (4 + n) r^2 
\nonumber\\
&-& 160 (a + b) \ddot\theta^3 \varepsilon r^3 + (a + b) \dot f^3 \dot\theta^9 \varepsilon r^9 - 
     2 (5 a + b) \dot f^2 \dot\theta^8 \varepsilon n r^7 \cot\theta 
\nonumber\\
&-&  8 \ddot\theta \dot\theta^2 \varepsilon r (b (240 + (47 - 16 n) n) + 
        a (240 + (68 - 13 n) n) + 36 (a + b) \ddot\theta n r^2 \cot\theta) 
\nonumber\\
&+&  8 \dot\theta^3 \varepsilon (b (-48 + 13 (-4 + n) n) + 
        a (2 + n) (-24 + n (-18 + 5 n)) + 30 (a + b) \ddot\theta^2 \dot f r^5 
\nonumber\\
&+&  2 \ddot\theta n (16 b n + a (-9 + 34 n)) r^2 \cot\theta) + 
     8 \dot\theta^4 r (\varepsilon n (b (-42 + 43 n) + a (-78 + n (76 + 15 n))) \cot\theta 
\nonumber\\
&-& \ddot\theta r^2 (16 (a + b) \varepsilon n + 3 r^2 + \varepsilon r (\dot f (-14 a - 23 b + 43 a n + 16 b n) 
+ 6 (3 a + 2 b) \ddot f r) 
\nonumber\\
&+&  \varepsilon (b (19 - 16 n) + a (16 - 13 n)) n \cot\theta^2)) + 
     2 \dot\theta^6 r^3 ((5 a + b) \ddot\theta \dot f^2 \varepsilon r^4 + 
        4 n \cot\theta (-7 b \varepsilon n + 5 a \varepsilon n^2 
\nonumber\\
&+&  4 r^2 + \varepsilon r (\dot f (2 b (7 + 2 n) + 7 a (1 + 6 n)) + (17 a + 
     4 b) \ddot f r) + \varepsilon (7 b + a (12 - 5 n)) (-2 + n) \csc\theta^2))
\nonumber\\
&+&  2 \dot f \dot\theta^7 r^5 (\varepsilon r (\dot f (-5 b (2 + n) - a (14 + 25 n)) - 
   2 (9 a + 5 b) \ddot f r) + 2 (\varepsilon n (b - 11 a n - 4 b n) - 5 r^2 
\nonumber\\
&+&  \varepsilon n (b (-5 + 4 n) + a (-12 + 11 n)) \csc\theta^2)) - 
     4 \dot\theta^5 r^2 (\varepsilon r (-2 \ddot f (-8 a + b + 19 a n + 4 b n) r 
\nonumber\\
&-&  4 (4 a + b) f^{(3)} r^2 +  \dot f (a (76 + (150 - 19 n) n) + b (58 + (9 - 4 n) n) + 
    6 (3 a + 4 b) \ddot\theta n r^2 \cot\theta)) 
\nonumber\\
&+&  2 (\varepsilon n (a (44 - 15 n) n + b (-6 + 17 n)) + (9 + 6 n) r^2 + \varepsilon n (b (20 - 17 n) 
\nonumber\\
&+&  a (50 + n (-62 + 15 n))) \csc\theta^2))) +  4 (-2 (10 a + 7 b) \dot f^3 \dot\theta^3 \varepsilon r^6 
+ \dot f^2 \varepsilon r^3 (2 (18 a + 17 b) \ddot\theta r 
\nonumber\\
&+&  \dot\theta (20 a (2 + n) +   6 b (6 + n) + (26 a + 7 b) \dot\theta n r \cot\theta)) + 
     2 n \cot\theta (-\varepsilon n (b + a n) + r^2 - 2 a \ddot f \varepsilon r^2 
\nonumber\\
&+&  \varepsilon (-2 + n) (b + a n) \csc\theta^2) + 
     2 \dot f r (2 \varepsilon n (2 a + b - 2 a n) \cot\theta 
\nonumber\\
&+&  \dot\theta r (-r^2 + 4 (2 a + b) \ddot f \varepsilon r^2 + 
           \varepsilon n ((b - 7 a n) \cot\theta^2 + 8 a \csc\theta^2))))\left.\right]\nonumber
\end{eqnarray}}
\begin{eqnarray}\label{s3}
s_3&=&\frac{1}{8(a+b)r^4f^2}\left[n(n-2)(1+r^2f\dot\theta^2)(b+an+2(b-a(n-2))r^2f\dot\theta^2) \right]\ ,\\
s_2&=&\frac{1}{16(a+b)r^3f^2}\left[
n(2\dot f\dot\theta(b+a(8-7n))r+4\dot\theta^5f^3(n-1)(a+b+an)r^4\right. 
\\\label{s2}
&+&f(2\dot\theta n(8a+b-7an)+2\ddot\theta(8a+b-7an)r+\dot f\dot\theta^3(7an+4bn-8a-5b)r^3) 
\nonumber\\
&+&
2\dot\theta^2f^2r^2(2\ddot\theta(b(4n-5)+a(5n-6))r+\dot\theta(b(7n-9)+a((27-5n)n-24) 
\nonumber\\
&+&\left. a\dot f\dot\theta^2(n-1)r^3)))\right]\ ,\nonumber\\
s_1&=&\frac{1}{32(a+b)r^4\varepsilon f^2F^2}\left[
n (-4 \varepsilon n (b + a n) + 4 r^2 + 4 \dot\theta^6 \varepsilon f^4 (-13 + 7 n) (a + b + a n) r^6 \right.
\\\label{s1}
&+& 2 \varepsilon r (-4 a \ddot f r + \dot f (8 a + 4 b - 8 a n + (26 a + 7 b) \dot f \dot\theta^2 r^3)) 
+ 4 \dot\theta^2 f^3 r^4 (4 \dot\theta^2 \varepsilon (b (-11 + 6 n) 
\nonumber\\
&+& a (-19 + n (8 + 3 n))) - 48 (a + b) \ddot\theta^2 \varepsilon r^2 + 
      2 \dot\theta \varepsilon r (\ddot\theta (-37 a - 25 b + 17 a n + 8 b n) 
\nonumber\\
&+& 8 (a + b) \theta^{(3)} r) + \dot\theta^4 r^2 (2 \varepsilon n (-b + a n) + r^2 + 
         \varepsilon r (3 \dot f (a + 2 b + 3 a n) + 4 a \ddot f r))) 
\nonumber\\
&+& f^2 r^2 (4 \dot\theta^2 \varepsilon (b (2 + 19 n) + a (-2 + n (44 + 3 n))) + 
      48 (a + b) \ddot\theta^2 \varepsilon r^2 
\nonumber\\
&-& 12 (3 a + 4 b) \ddot\theta \dot f \dot\theta^3 \varepsilon r^4 - (5 a + 
         b) \dot f^2 \dot\theta^6 \varepsilon r^6 + 
      8 \dot\theta \varepsilon r (\ddot\theta (28 a + 25 b + 17 a n + 8 b n) 
\nonumber\\
&+&  8 (a + b) \theta^{(3)} r) +  4 \dot\theta^4 r^2 (\varepsilon n (-5 b + 3 a n) + 3 r^2 + 
 \varepsilon r (\dot f (4 b (2 + n) + a (4 + 33 n)) 
\nonumber\\
&+& (13 a + 4 b) \ddot f r))) + 
   2 f (4 \varepsilon (b - a (-2 + n)) (-1 + n) + 
      \dot\theta r^2 (4 (28 a + 19 b) \ddot\theta \dot f \varepsilon r^2 
\nonumber\\
&+& \dot\theta (-8 b \varepsilon n + 6 r^2 + \varepsilon r (2 (7 a + 4 b) \ddot f r + 
\dot f (78 b + 8 b n + 20 a (7 + 2 n) 
\nonumber\\
&-& (9 a + 11 b) \dot f \dot\theta^2 r^3)))))) \left.\right]\nonumber\ ,\\
g(\dot\theta|_{r_1})&=&-((\dot\theta (2 (b + a (8 - 7 n)) (f n + \dot f r) + 
    2 \dot\theta^4 f^2 (-1 + n) r^4 (2 f (a + b + a n) + a \dot f r) \\
&+&  \dot\theta^2 f r^2 (-2 f (b (9 - 7 n) + a (24 + n (-27 + 5 n))) + 
       \dot f (b (-5 + 4 n) 
\nonumber\\
&+& a (-8 + 7 n)) r)))/(2 f r (b + a (8 - 7 n) + 2 \dot\theta^2 f (b (-5 + 4 n) 
\nonumber\\
&+& a (-6 + 5 n)) r^2)))\left.\right|_{r_1}\ ,\nonumber\\
h(\dot\theta|_{r_1},\ddot\theta|_{r_1})&=&
\frac{1}{64(a+b)\dot\theta\varepsilon f^2r^4 F^2}\left[\right.
(4 \varepsilon n (b + a n) -  4 r^2 \\
&-& 4 \dot\theta^6 \varepsilon f^4 (-13 + 7 n) (a + b + a n) r^6 + 
  2 \varepsilon r (4 a \ddot f r + \dot f (-4 b + 8 a (-1 + n) 
\nonumber\\
&-& (26 a + 7 b) \dot f \dot\theta^2 r^3)) - 
  4 \dot\theta^2 f^3 r^4 (4 \dot\theta^2 \varepsilon (b (-11 + 6 n) + a (-19 + n (8 + 3 n))) 
\nonumber\\
&+& 2 \ddot\theta \dot\theta \varepsilon (b (-25 + 8 n) + a (-37 + 17 n)) r - 
     48 (a + b) \ddot\theta^2 \varepsilon r^2 
\nonumber\\
&+&  \dot\theta^4 r^2 (2 \varepsilon n (-b + a n) + r^2 + 
        \varepsilon r (3 \dot f (a + 2 b + 3 a n) + 4 a \ddot f r))) 
\nonumber\\
&+&  2 f (-4 \varepsilon (b - a (-2 + n)) (-1 + n) - 
     4 (28 a + 19 b) \ddot\theta \dot f \dot\theta \varepsilon r^4 
 \nonumber\\
&+& (9 a + 11 b) \dot f^2 \dot\theta^4 \varepsilon r^6 - 
     2 \dot\theta^2 r^2 (-4 b \varepsilon n + 3 r^2 + 
        \varepsilon r (\dot f (70 a + 39 b + 4 (5 a + b) n) 
\nonumber\\
&+& (7 a + 4 b) \ddot f r))) + f^2 r^2 (-4 \dot\theta^2 \varepsilon (b (2 + 19 n) 
+ a (-2 + n (44 + 3 n))) 
\nonumber\\
&-&  8 \ddot\theta \dot\theta \varepsilon (b (25 + 8 n) + a (28 + 17 n)) r - 
48 (a + b) \ddot\theta^2 \varepsilon r^2 + 12 (3 a + 4 b) \ddot\theta \dot f \dot\theta^3 \varepsilon r^4 
\nonumber\\
&+&(5 a + b) \dot f^2 \dot\theta^6 \varepsilon r^6 - 
     4 \dot\theta^4 r^2 (\varepsilon n (-5 b + 3 a n) + 3 r^2 + 
        \varepsilon r (\dot f (4 b (2 + n) 
\nonumber\\
&+& a (4 + 33 n)) + (13 a + 4 b) \ddot f r))))\left.\right]\left.\right|_{r_1}\ ,\nonumber\\
y_2&=&\frac{1}{16 (a + b) \varepsilon r^4}\left[\right.
-2 (10 a + 7 b) \dot f^3 \dot\theta^3 \varepsilon r^6 + 
 \dot f^2 \varepsilon r^3 (2 (18 a + 17 b) \ddot\theta r + 
    \dot\theta (20 a (2 + n) \\
&+&  6 b (6 + n) + (26 a + 7 b) \dot\theta n r \cot\theta)) + 
 2 n \cot\theta (r^2 - 2 a \ddot f \varepsilon r^2 
 \nonumber\\
&+&   \varepsilon (b + a n) (-n + (-2 + n) \csc^2\theta)) + 
 2 \dot f r (2 \varepsilon (b - 2 a (-1 + n)) n \cot\theta 
\nonumber\\
&+& \dot\theta r (-r^2 + 4 (2 a + b) \ddot f \varepsilon r^2 + 
       \varepsilon n ((b - 7 a n) \cot^2\theta + 8 a \csc^2\theta)))
\left.\right]\left.\right|_{r_0}\ ,\nonumber
\end{eqnarray}
\begin{eqnarray}
y_1&=&\frac{1}{32 (a + b) \varepsilon r^4}\left[\right.
(98 a + 83 b) \dot f^3 \dot\theta^5 \varepsilon r^8 - 
 2 (35 a + 18 b) \dot f^2 \dot\theta^4 \varepsilon n r^6 \cot\theta \\
&+& 4 \dot\theta r (\varepsilon n^2 (-b + 7 a n) + (-2 - n) r^2 + 
    2 \varepsilon r (r (2 \ddot f (3 a (2 + n) + b (5 + n)) 
\nonumber\\
&+& (2 a + b) f^{(3)} r) + \dot f (2 a (2 + n) (2 + 3 n) +  b (8 + n (15 + n)) 
\nonumber\\
&+& (28 a + 19 b) \ddot\theta n r^2 \cot\theta)) 
+ \varepsilon (b + a (8 - 7 n)) n^2 \csc^2\theta) 
\nonumber\\
&+& 2 \dot f \dot\theta^3 r^4 (\varepsilon n (b - 7 a n - 4 b n) - 3 r^2 + 
    \varepsilon r (\dot f (-6 a (34 + 7 n) - b (174 + 19 n)) 
\nonumber\\
&-&       4 (8 a + 7 b) \ddot f r) + 
    \varepsilon n (b (-5 + 4 n) + a (-8 + 7 n)) \csc^2\theta) 
\nonumber\\
&+& 2 \dot\theta^2 r^2 (-(210 a + 197 b) \ddot\theta \dot f^2 \varepsilon r^4 + 
    2 n \cot\theta (\varepsilon n (-3 b + a n) + 2 r^2 
 \nonumber\\
&+&       \varepsilon r (\dot f (66 a + 37 b + 4 (6 a + b) n) + (9 a + 4 b) \ddot f r) 
 \nonumber\\
&+&       \varepsilon (3 b - a (-4 + n)) (-2 + n) \csc^2\theta)) + 
 4 (32 (a + b) \theta^{(3)} \dot f \varepsilon r^4 
 \nonumber\\
&+&    2 \varepsilon (b - a (-2 + n)) (-1 + n) n \cot\theta + 
    \ddot\theta r^2 (-r^2 + 
       2 \varepsilon r (\dot f (20 a (2 + n) 
\nonumber\\
&+& b (40 + 17 n)) + (10 a + 9 b) \ddot f r) + 
       \varepsilon n ((b - 7 a n) \cot^2\theta + 8 a \csc^2\theta)))
\left.\right]\left.\right|_{r_0}\ ,\nonumber\\
G(\dot\theta|_{r_0})&=&\frac{1}{2 (18 a + 17 b) \dot f^2\varepsilon  r^4}\left[\right.
2 (10 a + 7 b) \dot f^3 \dot\theta^3 \varepsilon r^6 - 
 \dot f^2 \dot\theta \varepsilon r^3 (20 a (2 + n) + 
    6 b (6 + n) 
\\
&+& (26 a + 7 b) \dot\theta n r \cot\theta) + 
 2 n \cot\theta (\varepsilon n (b + a n) - r^2 + 2 a \ddot f \varepsilon r^2 - 
    \varepsilon (-2 + n) (b + a n) \csc^2\theta) 
 \nonumber\\
&+& 2 \dot f r (-2 \varepsilon n (2 a + b - 2 a n) \cot\theta + 
    \dot\theta r (\varepsilon n (b - 7 a n) + r^2 - 4 (2 a + b) \ddot f \varepsilon r^2 
 \nonumber\\
&+&       \varepsilon n (-8 a - b + 7 a n) \csc^2\theta))
\left.\right]\left.\right|_{r_0}\ ,\nonumber\\
H(\dot\theta|_{r_0},\ddot\theta|_{r_0})&=&
-\frac{1}{128 (a + b) \dot f r^4}\left[\right.
8 (b - a (-2 + n)) (-1 + n) n \cot\theta + (1/\varepsilon)
 r ((98 a + 83 b) \dot f^3 \dot\theta^5 \varepsilon r^7 
\\
&-&    2 (35 a + 18 b) \dot f^2 \dot\theta^4 \varepsilon n r^5 \cot\theta + 
    4 \ddot\theta r (\varepsilon n (-b + 7 a n) - r^2 + 
       2 \varepsilon r (\dot f (20 a (2 + n) 
\nonumber\\
&+& b (40 + 17 n)) + (10 a + 
             9 b) \ddot f r) + \varepsilon (b + a (8 - 7 n)) n \csc^2\theta) + 
    4 \dot\theta (\varepsilon n^2 (-b + 7 a n) 
\nonumber\\
&+& (-2 - n) r^2 +  2 \varepsilon r (r (2 \ddot f (3 a (2 + n) + b (5 + n)) + (2 a + b) f^{(3)} r) 
+  \dot f (2 a (2 + n) (2 + 3 n) 
\nonumber\\
&+& b (8 + n (15 + n)) + (28 a + 19 b) \ddot\theta n r^2 \cot\theta)) 
+ \varepsilon (b + a (8 - 7 n)) n^2 \csc^2\theta) 
 \nonumber\\
&+&    2 \dot f \dot\theta^3 r^3 (\varepsilon n (b - 7 a n - 4 b n) - 3 r^2 + 
       \varepsilon r (\dot f (-6 a (34 + 7 n) - b (174 + 19 n)) 
 \nonumber\\
&-&  4 (8 a + 7 b) \ddot f r) + 
       \varepsilon n (b (-5 + 4 n) + a (-8 + 7 n)) \csc^2\theta) + 
    2 \dot\theta^2 r (-(210 a + 197 b) \ddot\theta \dot f^2 \varepsilon r^4 
 \nonumber\\
&+& 2 n \cot\theta (\varepsilon n (-3 b + a n) + 2 r^2 + 
          \varepsilon r (\dot f (66 a + 37 b + 4 (6 a + b) n) + (9 a + 4 b) \ddot f r) 
\nonumber\\
&+& \varepsilon (3 b - a (-4 + n)) (-2 + n) \csc^2\theta)))
\left.\right]\left.\right|_{r_0}\ . \nonumber
\end{eqnarray}

\end{widetext}

\end{document}